\documentclass[12pt,preprint]{aastex}

\slugcomment{
}

\newcommand{\beq}{\begin{equation}}
\newcommand{\beqa}{\begin{eqnarray}}
\newcommand{\eeq}{\end{equation}}
\newcommand{\eeqa}{\end{eqnarray}}

\newcommand{\siml}{\lesssim}

\newcommand{\bm}[1]{\mbox{{\boldmath $#1$}}} 

\begin{document}


\title{  
Delayed Flashes from Counter Jets of Gamma Ray Bursts
}


\author{Ryo Yamazaki\altaffilmark{1}}
\affil{Department of Physics, Kyoto University, Kyoto 606-8502, Japan}
\email{yamazaki@tap.scphys.kyoto-u.ac.jp}

\and

\author{Kunihito Ioka\altaffilmark{2}}
\affil{Department of Earth and Space Science, Osaka University, 
Toyonaka 560-0043, Japan}
\email{ioka@vega.ess.sci.osaka-u.ac.jp}

\and

\author{Takashi Nakamura\altaffilmark{3}}
\affil{Department of Physics, Kyoto University, Kyoto 606-8502, Japan}
\email{takashi@tap.scphys.kyoto-u.ac.jp}

%
%
%


\def\d{{\rm d}}
\def\p{\partial}
\def\w{\wedge}
\def\o{\otimes}
\def\f{\frac}
\def\tr{{\rm tr}}
\def\Half{\frac{1}{2}}
\def\half{{\scriptstyle \frac{1}{2}}}
\def\T{\tilde}
\def\RA{\rightarrow}
\def\N{\nonumber}
\def\n{\nabla}
\def\bb{\bibitem}
\def\BE{\begin{equation}}
\def\EE{\end{equation}}
\def\BEA{\begin{eqnarray}}
\def\EEA{\end{eqnarray}}
\def\L{\label}
\def\zero{{\scriptscriptstyle 0}}
\begin{abstract}
If X-ray flashes are due to the forward jet emissions from
gamma ray bursts (GRBs) observed with large viewing angles, 
we show that a prompt emission from a counter jet
should be observed as a {\it delayed flash} in the UV or optical band 
several hours to a day after the X-ray flash.
Ultraviolet and Optical Telescope on {\it Swift} can observe 
the delayed flashes within $\sim13$ Mpc, 
so that (double-sided) jets
of GRBs may be directly confirmed.
Since the event rate of delayed flashes
detected by {\it Swift} may be as small as 
$\sim6\times10^{-5}$events yr$^{-1}$,
we require more sensitive detectors in future experiments.

\end{abstract}


\keywords{gamma rays: bursts ---gamma rays: theory}


\clearpage
\section{INTRODUCTION}
Several observations suggest that gamma-ray bursts (GRBs) 
are caused by relativistic jets (e.g., Frail et al. 2001).
However, in order to establish 
%
that GRBs are collimated,
other observations are indispensable, such as
polarization observations (Ghisellini \& Lazzati 1999; Sari 1999)
and microlensing observations (Ioka \& Nakamura 2001b).
Some theoretical models of jet emissions have been discussed
(Totani \& Panaitescu 2002; Huang, Dai \& Lu 2002; 
Dado, Dar \& De~R\'{u}jula 2001).
If GRBs are due to forward jet emissions,
there should most likely be counter jet emissions,
as in the AGN (Begelman, Blandford \& Rees 1984)
and the microquasar (Mirabel \& Rodr\'{\i}guez 1999).
Therefore the detection of counter jet emissions will give us
direct evidence for the jet model of GRBs.

The confirmation of a counter 
jet has been by far the most important factor in the jet model of
astrophysical objects. A mysterious spot was found in SN1987A 
using the speckle technique
(Meikle et al. 1987; Nisenson et al. 1987). 
Many models including the jet model were proposed
(Rees 1987; Piran \& Nakamura 1987).  
At that time, it was difficult to distinguish each model from observations 
since only one spot was found. 
In the jet model, the counter jet should be observed although 
it is dim due to redshifting (Piran \& Nakamura 1987). 
However, later in 1999, two spots were confirmed using new software
to analyze the speckle data (Nisenson \& Papaliolios 1999).
Very recently the jet feature of the ejecta of SN1987A 
whose position angle is the same as the mysterious spot was confirmed 
by the HST (Wang et al. 2002).  As a result the jet model 
by Piran \& Nakamura (1987) took the advantage.
Furthermore the observation of a counter jet may enable
us to estimate the Lorentz factor of the jet,
as for the AGN and microquasar. 
Therefore it is important to argue the observational properties 
of the emission from the counter jet of a GRB.

Let us consider the emission from a counter jet
with a Lorentz factor $\gamma$.
The observed typical frequency of the counter jet emission
is about $\gamma^2$ times smaller than that of the 
forward jet (i.e., the GRB).
Since the typical frequency of the GRB is $\sim100$ keV,
the typical observed frequency of the counter jet emission is
$\sim10(\gamma/100)^{-2}$ eV,
which is in the UV or optical band.
This transient phenomenon
should be observed about several tens of hours after
the forward jet emission,
%
since it is at a radius of order $10^{14}$--$10^{15}$ cm
that photons are emitted from each jet leaving, almost simultaneously,
the central engine.
We call this event the {\it delayed flash} (DF).

Any attempt to detect the DF might be difficult since
the afterglow of the forward jet might be brighter than the DF.
The afterglow of the GRB, i.e., the afterglow of the on-axis forward jet,
is much brighter than the DF.
However if the forward jet is observed with a large viewing angle,
there is a chance to observe the DF since the forward jet emission
is also dim at the time of the DF.

Recently, we studied the emission from the off-axis jet
(Yamazaki, Ioka, \& Nakamura 2002a, b; see also Nakamura 2000;
Ioka \& Nakamura 2001a).
We proposed that if we observe a GRB with a large viewing angle,
it looks like an X-ray flash (XRF),
a new class of X-ray transients which has been 
recently recognized as a phenomenon related to the GRB
(e.g., Heise et al. 2001; Kippen et al. 2002;
Barraud et al. 2002).
The off-axis jet model can explain the typical observed frequency 
and other observational characteristics of the XRF, such as
the peak flux ratio and the fluence ratio between the $\gamma$-ray
and the X-ray band, the X-ray photon index, the typical duration,
and the event rate (Yamazaki, Ioka, \& Nakamura 2002a, b).
Although the origin of XRFs is uncertain,
we assume that XRFs arise from prompt off-axis 
jet emissions hereafter.

In this paper, we will show that
the DF can be observed after the XRF in principle.
We will calculate the light curves of
the XRF, DF, and the afterglow of the XRF,
and discuss whether the DF can be detected by 
the {\it Swift} satellite.
We will find that we need more sensitive detectors
to detect the DF.
In \S~\ref{sec:model}, we describe a simple forward-/counter-jet 
model for the XRF and DF.
In \S~\ref{sec:lightcurve} and \S~\ref{sec:afterglow},
we show the light curves of the XRF, the DF,
and the XRF afterglow.
\S~\ref{sec:dis} is devoted to a discussion.

\section{INSTANTANEOUS EMISSION FROM AN EXPANDING JET}\label{sec:model}
We extend the simple jet model 
by Ioka \& Nakamura (2001) and Yamazaki, Ioka, \& Nakamura (2002a).
In these works we assume 
that the shell width $l$ is much smaller than
the separation of the shells $L$ in the internal shock,
since the separation $L$ mainly determines the emission timescale
of the forward jet.
However in this paper we consider a finite shell width, since
the shell width $l$ determines the emission timescale
of the counter jet.
The cooling timescale is much shorter than other
timescales (Sari, Narayan \& Piran 1996),
so we assume an instantaneous emission at the shock 
front as before.
We use a spherical coordinate system $(t,r,\theta,\phi)$
in the Lab frame, where the $\theta=0$ axis points toward the detector,
and the central engine is located at $r=0$.
The forward jet has a viewing angle, $\theta_v>0$,
which the axis of the emission cone makes with the 
$\theta=0$ axis, while the counter jet has a viewing angle
$\theta_v+\pi$.
When the emitting shock front moves radially from $t=t_0$ and $r=r_0$
with the Lorentz factor $\gamma=1/\sqrt{1-\beta^2}$,
the emissivity for the XRF has a functional form of
\beqa
j'_{\nu'}(\Omega_d', {\bm r}, t)&=&A(t) f(\nu')
\delta[r-r_0-\beta c(t-t_0)]
\nonumber\\
&\times& H(\Delta \theta-|\theta-\theta_v|)
H\left[\cos \phi-\left({{\cos\Delta \theta-\cos\theta_v 
\cos\theta}\over
{\sin \theta_v \sin\theta}}\right)\right],
\eeqa
where $f(\nu')$ represents the spectral shape.
The Heaviside step function $H(x)$ describes that the emission is 
inside a cone of an opening half-angle $\Delta \theta$.
Then the observed flux per unit frequency 
of a single pulse at the observed time $T$
is given by
\beqa
F_{\nu}(T)={{2 {r_0}^{2} \gamma^2}\over{\beta D^2 (r_0/c\beta)}}
\int dt A(t) 
{{[\gamma^2(1-\beta \cos\theta(T))]}\over
{[\gamma^2(1-\beta \cos\theta(t))]}}
{{\Delta \phi(t) f[\nu \gamma(1-\beta \cos \theta(t))]}\over
{[\gamma^2(1-\beta \cos\theta(t))]^2}},
\label{eq:jetthin}
\eeqa
where
$1-\beta\cos\theta(T)=(c\beta/r_0)(T-T_0)$,
$1-\beta\cos\theta(t)=[1-\beta\cos\theta(T)]/
[(c\beta/r_0)(t-T_0)]$ and $T_0=t_0-r_0/c\beta$
(see Ioka \& Nakamura 2001a).
For the XRF, 
$\theta(t)$ varies from ${\rm max}\{0,\theta_v-\Delta\theta\}$
to $\theta_v+\Delta \theta$.
The function $\Delta \phi(t)$ is given as
\begin{equation}
\Delta\phi(t)=
\left\{
\begin{array}
{c@{} l@{}}
\pi & (\theta_v<\Delta\theta \>\>{\rm and}\>\> 
0<\theta(t)\leq \Delta\theta-\theta_v )\\
\cos^{-1}\left[
\f{\cos \Delta \theta - \cos \theta(t) \cos \theta_v}
{\sin \theta_v \sin \theta(t)}\right]
& ({\rm otherwise}) 
\end{array} \right. .
\end{equation}
Also for the DF we can use equation (\ref{eq:jetthin}).
For the DF, 
$\theta(t)$ varies from $\pi+{\rm max}\{0,\theta_v-\Delta\theta\}$
to $\pi+\theta_v+\Delta \theta$,
and the function $\Delta \phi(t)$ is given as
\begin{equation}
\Delta\phi(t)=
\left\{
\begin{array}
{c@{} l@{}}
\pi & (\theta_v<\Delta\theta \>\>{\rm and}\>\> 
\pi<\theta(t)\leq \pi+\Delta\theta-\theta_v )\\
\cos^{-1}\left[
\f{\cos \Delta \theta + \cos \theta(t) \cos \theta_v}
{-\sin \theta_v \sin \theta(t)}\right]
& ({\rm otherwise}) 
\end{array} \right. .
\end{equation}

The normalization of emissivity $A(t)$ 
is determined by the hydrodynamics.
Here for simplicity we adopt the following functional form,
\beqa
A(t)=A_0\left(\frac{t-T_0}{r_0/c\beta}\right)^{-2}
H(t-t_0) H(t_e-t),
\label{eq:At}
\eeqa
where the emission ends at $t=t_e$ and
the released energy at each distance $r$ is constant.
Our conclusion does not depend on $t_e$ or the functional form
so much.
A pulse-starting time and ending time are given as
\begin{eqnarray}
T_{start}^{(XRF)}&=&T_0+({r_0}/{c\beta})
(1-\beta\cos(\max[0,\theta_v-\Delta\theta])),
\label{eq:tstart:xrf}
\\
T_{end}^{(XRF)}&=&T_0+[({r_0}/{c\beta})+t_e-t_0]
(1-\beta\cos(\theta_v+\Delta\theta)),
\label{eq:tend:xrf}
\end{eqnarray}
for the XRF, and 
\begin{eqnarray}
T_{start}^{(DF)}&=&T_0+({r_0}/{c\beta})
(1+\beta\cos(\theta_v+\Delta\theta)),
\label{eq:tstart:uvf}
\\
T_{end}^{(DF)}&=&T_0+[({r_0}/{c\beta})+t_e-t_0]
(1+\beta\cos(\max[0,\theta_v-\Delta\theta])),
\label{eq:tend:uvf}
\end{eqnarray}
for the DF.

The spectrum of the GRB is well approximated by 
the Band spectrum (Band et al. 1993). 
In order to have a spectral shape similar to the Band spectrum,
we adopt the following form of the spectrum in the comoving frame,
\beqa
f(\nu')=(\nu'/\nu'_0)^{1+\alpha_B}
(1+\nu'/\nu'_0)^{\beta_B-\alpha_B},
\label{eq:spectrum}
\eeqa
where $\alpha_B$ ($\beta_B$) is the low (high) energy power law index.
In the GRB, $\alpha_B\sim -1$ and $\beta_B\sim -3$ are typical values
(Preece et al. 2000).
Equations (\ref{eq:jetthin}), (\ref{eq:At}) and (\ref{eq:spectrum})
are the basic equations to calculate the flux of a single pulse,
which depends on following parameters:
$\gamma \gg 1$, $\theta_v\ll 1$, $\Delta \theta \ll 1$,
$\gamma \nu_0'$, $r_0/c \beta \gamma^2$, $\alpha_B$, $\beta_B$, 
$D$, $A_0$, $t_0$ and $t_e$.

Hereafter we consider mainly the following canonical set of parameters;
$\gamma=100$, $\gamma\Delta\theta=5$, 
$r_0/\beta c\gamma^2=1\,{\rm s}$, $\alpha_B=-1$, $\beta_B=-3$,
$h\gamma\nu'_0=200\,{\rm keV}$ and $t_0=r_0/c\beta$.
We adopt $t_e=1.3 r_0/c\beta$
since most pulses rise more quickly than they decay
(Norris et al. 1996).
The value $\gamma\Delta\theta=5$ has been obtained from
the fitting of the afterglow light curve
(Frail et al. 2001; see also Panaitescu \& Kumar 2002).
We fix the amplitude $A_0$ so that
the isotropic $\gamma$-ray energy 
$E_{iso}=4\pi D^2 S(20-2000\,{\rm keV})$
satisfies
$(\Delta\theta)^2E_{iso}=1\times10^{51}{\rm ergs}$
when $\gamma\theta_v=0$ (Frail et al. 2001).
Here $S(\nu_1-\nu_2)=
\int_{T_{start}}^{T_{end}}F(T;\nu_1-\nu_2)dT$
is the observed fluence in the energy range $\nu_1-\nu_2$
and $F(T;\nu_1-\nu_2)=\int_{\nu_1}^{\nu_2}
F_\nu(T)d\nu$ is the observed flux in the same energy range.
Then, we obtain $A_0=0.24\,{\rm erg}\ {\rm s}^{-1}
\ {\rm cm}^{-2} \ {\rm Hz}^{-1}$ for the fiducial parameters.
Note that the observed flux is proportional to $D^{-2}$.\footnote{
When we consider the effect of cosmology ($\Omega_M=0.3$,
$\Omega_\Lambda=0.7$, and $h=0.7$), $D\sim1\,{\rm Gpc}$ corresponds
to $z\sim0.2$. 
Since we consider the case $D<1\,{\rm Gpc}$ in the following sections,
this does not affect our argument qualitatively but alters
the quantitative results up to a factor of $2$.
} 

\section{LIGHT CURVES OF X-RAY FLASH AND DELAYED FLASH}
\label{sec:lightcurve}
In this section,
we plot the light curves of the XRF and the DF
using Eq.(\ref{eq:jetthin}), and
discuss whether these events can be observed by 
{\it Swift} satellite.

First, we show the light curves of the XRF
$F(T;\,15-150\,{\rm keV})$ in Figure 1 with varying $\gamma\theta_v$.
The observation band corresponds to that of 
the Burst Alert Telescope (BAT) on {\it Swift}.
As $\gamma\theta_v$ increases, the peak flux of the XRF
$F_{peak}^{(XRF)}$ decreases due to the relativistic beaming effect.

The light curves of the DF $F(T;\,1.9-7.3\,{\rm eV})$ are shown
in Figure 2 with varying $\gamma\theta_v$.
The observation band corresponds that of 
the Ultraviolet and Optical Telescope (UVOT) on {\it Swift}.
We find that the flux remains almost constant in each pulse,
and that the peak flux $F_{peak}^{(DF)}$ does not depend on
the viewing angle $\gamma\theta_v$ so much.
This is because
the value of $\theta(t)$ ranges between 
$\pi+{\rm max}\{0,\theta_v-\Delta\theta\}$ and
$\pi+\theta_v+\Delta\theta$
and $(c\beta/r_0)(t-T_0)\sim 1$
so that
$\theta(t)\sim \theta(T)\sim \pi$ in equation (\ref{eq:jetthin}).
From equation (\ref{eq:jetthin}) the peak flux of the DF
can be estimated as
$F_{peak}^{(DF)}
\sim \Delta \nu F_{\nu}
\sim \Delta \nu (2r_0^2 \gamma^2/\beta D^2) A_0 \Delta \phi f/(2\gamma^2)^2
\sim 10^{-19}$ ergs s$^{-1}$ cm$^{-2}$,
where
$\Delta \nu\sim 10^{14}$ Hz,
$\Delta \phi\sim \Delta \theta/\theta_v\sim 0.1$ and
$f\sim 0.2$.


The limiting sensitivity of the UVOT (BAT) can be estimated as
$1\times10^{-15}\, {\rm ergs}\ \ {\rm s}^{-1}\>{\rm cm}^{-2}$
($5\times10^{-10}\, {\rm ergs}\ \ {\rm s}^{-1}\>{\rm cm}^{-2}$)
for a duration of $\sim5\times10^3$ ($\sim10^2$) seconds.
The BAT localizes the XRF and the following observation by the UVOT
may identify  the associated DF.
One can see that the DF with $D\lesssim 13$ Mpc is observable.
Then the BAT can detect the preceding XRF if
$\gamma\theta_v\lesssim 30$.

\section{AFTERGLOW OF X-RAY FLASH}\label{sec:afterglow}
The start and end time of the DF is about
$T_{start}^{(DF)}\sim 2 t_0 \sim 2\times 10^4$ s
and $T_{end}^{(DF)} \sim 2 t_e \sim 2.6 \times 10^4$ s
for $\gamma=10^2$, $t_0=r_0/c\beta=10^4$ s
and $t_e=1.3 t_0$.
Therefore, one should compare the flux of the DF
with that of the XRF afterglow.
In this section, we plot the light curves of the XRF
afterglow and see whether or not the DF can be detected.
We use model 1 of Granot et al. (2002) as a
simple model of the off-axis afterglow emission from
the collimated jet (see also Dalal, Griest, \& Pruet 2002).

For $\theta_v=0$, the standard afterglow model, i.e., 
the synchrotron-shock model,
can explain observational properties of the GRB afterglow very well 
(Piran, 1999), and gives the observed flux per unit frequency as
$F_\nu(T;\,\theta_v=0)=F_\nu^{({\rm R-SPH})}\equiv G(\nu,T)$,
where $F_\nu^{({\rm R-SPH})}$ is the observed flux given by 
Rhoads (1999) and Sari, Piran \& Halpern (1999).
For $\theta_v>\Delta\theta$, 
the emission is assumed to be
from a point source moving along the jet axis.
Then the flux is given by
$F_\nu(T;\,\theta_v)=a^3G(a^{-1}\nu,aT)$,
where $a\equiv(1-\beta)/(1-\beta\cos\tilde{\theta})\sim
(1+(\gamma\tilde{\theta})^2)^{-1}$.
We choose $\tilde{\theta}={\rm max}(0,\theta_v-\Delta\theta)$
to make this simple model more realistic (Granot et al. 2002). 
The Lorentz factor of the shell $\gamma$ can be determined by
\begin{equation}
\gamma\Delta\theta=
\left\{
\begin{array}
{c@{} l@{}}
(aT/t_{jet})^{-3/8} & \quad {\rm if} \quad aT<t_{jet} \\
(aT/t_{jet})^{-1/2} & \quad {\rm if} \quad aT>t_{jet},
\end{array} \right. 
\label{JetBreak}
\end{equation}
where $t_{jet}=1.9\times10^4 {\rm sec}\> n^{-1/3}
(\Delta\theta/0.05)^{8/3}(E/2\times10^{54} \,{\rm ergs})^{1/3}$ 
is the jet-break time observed from an on-axis observer
\footnote{For simplicity, we assume the relation $t'=R/(4\gamma^2c)$,
where $t'$ and $R$ are the time measured by an on-axis observer and
the radius of the shock.},
where $E$ is the isotropic equivalent value of the 
total energy in the shock.
We assume $E=\eta_\gamma^{-1} E_{iso}$ with a constant factor
$\eta_\gamma=0.2$, which is adopted in Frail et al. (2001).
Then, we obtain $E=2\times10^{54}\,{\rm ergs}$ and 
the geometry corrected total energy in the shock 
$(\Delta\theta)^2E/2=2.5\times10^{51}$~ergs.

Using above equations, we calculate the light curves of the afterglow.
In order to study the dependence on the viewing angle $\theta_v$,
we fix the rest of the parameters:
the power-law index of accelerated electrons $p=2.25$,
the number density of the ambient matter $n=1\,{\rm cm}^{-3}$,
$\varepsilon_e=0.1$ and $\varepsilon_B=0.01$,
and the distance $D=1\,{\rm Gpc}$.
Figure~3 (Figure~4) shows the result in the case of
$\Delta\theta=0.05$ ($\Delta\theta=0.1$).
The observation band is 1.9--7.3 eV, which corresponds to that of
UVOT.

We also plot the UV flux of the DF in the same figures.
We can see that for the canonical set of parameters
($\Delta\theta=0.05$, $\gamma=100$, and $r_0/c\beta\gamma^2=1$~sec), 
the UV flux of the DF dominates the afterglow
when $\theta_v\gtrsim0.21$.
For comparison, we show the light curves of the DF with one of
parameters changed from the fiducial value.
For large $\gamma$,
it is difficult to detect the DF since
the starting and ending time of the DF is late and the flux of
the DF is low due to the strong beaming effect.
When we alter $r_0/c\beta\gamma^2$,
the starting (and ending) time and the flux of the DF have a 
dependence $\propto (r_0/c\beta\gamma^2)$ and 
$\propto (r_0/c\beta\gamma^2)^{-1}$, respectively.
So the large $r_0/c\beta\gamma^2$ case has qualitatively
the same behavior as the large $\gamma$ case.
One can easily find that when $\Delta\theta$ becomes large, 
the flux of the afterglow of the XRF becomes
large while the light curves of the DF remains almost unchanged.
Therefore, we can conclude that the DF from a jet with smaller 
$\Delta\theta$, $\gamma$ and $r_0/c\beta\gamma^2$ 
have larger chance to be seen.
In consequence, according to the off-axis jet model,
it is preferable for the detection of the DF
that the preceding XRF has a low peak energy of a few keV,
a small variability owing to large $\theta_v$ 
(Yamazaki, Ioka, \& Nakamura 2002a,b) and a short
duration due to small $r_0/c\beta\gamma^2$.

We have used in this section model 1 of Granot et al. (2002).
A more realistic model for the off-axis emission from the
forward jet, model 3 of Granot et al. (2002), 
may have a more moderate
rise before the peak of the observed light curve
than the model we have adopted.
However, we consider the case in which the viewing angle is as large
as $\theta_v\gtrsim 5\Delta\theta$,
so that the differences between  these models may be small.

\section{DISCUSSION}\label{sec:dis}

We have calculated the light curves of the DF, XRF,
and the afterglow of the XRF.
We have shown that in principle,
the DF emission can be seen in the UV band about $10^4$--$10^5$ 
seconds after the XRF if the viewing angle is large enough 
(about 0.2--0.3 rad) for the 
afterglow of the XRF to be dimmer than the DF.
Since the UV flux of the GRB afterglow is much larger than
that of the DF, only the DF associated with an
off-axis jet, i.e., an XRF has any chance to be observed.
The preceding XRF should have a low peak energy of
a few keV, a small variability, and a short duration
for the DF to be detected.
Due to the relativistic beaming effect, the flux of the DF
is so small that only nearby events ($\lesssim13$~Mpc for the 
canonical parameters) can be observed by UVOT on {\it Swift}.
Following Yamazaki, Ioka, \& Nakamura (2002a),
we can roughly estimate the event rate of the DF 
for the instruments on {\it Swift} as
$R_{DF}\sim 6\times10^{-5}\,{\rm events}\ {\rm yr}^{-1}$,
where we adopt 
the event rate of the GRBs 
$r_{GRB}={5\times 10^{-8}\,{\rm events}\ {\rm yr}^{-1}\ 
{\rm galaxy}^{-1}}$
and the number density of galaxies
$n_g=10^{-2}\,{\rm galaxies}\ {\rm Mpc}^{-3}$.
%
%
Therefore,
we need next-generation detectors,
which are more sensitive than the instruments on {\it Swift},
to detect the DF associated with very dim XRF more frequently.

The DF may be obscured by dust extinction.
In fact, about half of accurately localized GRBs do not produce
a detectable optical afterglow 
(Fynbo et al. 2001; Lazzati, Covino, \& Ghisellini 2002).
One explanation for these ``dark GRBs''
is that most GRBs occur in giant molecular clouds
(e.g., Reichart \& Price 2002).
In this picture, a GRB has a detectable optical afterglow
only if the burst and the afterglow destroy the dust along the line of sight
to the observer
(Waxman \& Draine 2000; Fruchter, Krolik, \& Rhoads 2001),
as suggested by the comparison between X-ray and optical extinction
(Galama \& Wijers 2001).
In this case the DF is obscured since
the flux of the XRF
is too dim to carve out a path for the DF.
However this picture may have some problems, 
such as no evidence of an ionized absorber (Piro et al. 2002)
and variable column density in the X-ray afterglow 
(Djorgovski et al. 2001c).
There are other explanations for dark GRBs,
such as high redshift effects,
dust extinction in the interstellar medium of the host galaxy
(Ramirez-Ruiz, Trentham, \& Blain 2002; Piro et al. 2002)
and so on.
Therefore at present we cannot conclude that the DF is obscured.

If we assume that the absolute magnitude of the host galaxy is about
$\sim -20$ mag (Djorgovski et al. 2001a, b),
the apparent magnitude is about $\sim 20 + 5 \log D_{\rm Gpc}$.
Since a host galaxy with a size $\sim 10$ kpc has 
an angular size of $\sim 10 D_{\rm Gpc}^{-1}$ arcsec,
we can observe a point source which is dimmer than the host galaxy
by $\sim 10^{-4} D_{\rm Gpc}^{2}$
if the angular resolution is $\sim 0.1$ arcsec.
Therefore the DF has to be brighter than $\sim 30$ mag,
and we can observe the DF if $D \siml 13$ Mpc.

If the GRB is associated with a supernova (SN),
the emission from the SN may hide the DF.
The UV flux of SN1998bw was about $\sim 17$ mag at 
the distance $D \sim 40$ Mpc (Galama et al. 1998), i.e., 
$\sim 6\times 10^{-15} D_{\rm Gpc}^{-2}$ erg s$^{-1}$ cm$^{-2}$, 
so that a SN like SN1998bw is brighter than the DF.
However at present it is not clear whether all GRBs are 
associated with the SNe or not (e.g., Price et al. 2002).
In any cases, deep searches following the XRF will 
give us valuable information.

If the DF associated with an XRF is observed,
we will be able to estimate the Lorentz factor 
and the viewing angle of the jets.
Let the typical frequency or the break energy of the DF (XRF) be
$\nu_{DF}=\delta_{DF}\nu'_0$ ($\nu_{XRF}=\delta_{XRF}\nu'_0$),
where
$\delta\equiv 1/\gamma(1-\beta\cos\theta_v)$ is 
the Doppler factor.
When $\theta_v\ll1$, $\gamma\gg1$ and $(\gamma\theta_v)^2\gg1$,
we can derive $\delta_{DF}\sim1/(2\gamma)$ and
$\delta_{XRF}\sim2\gamma/(\gamma\theta_v)^2$.
Since we assume that the XRF is the GRB observed from
the off-axis viewing angle,
we may use the observational consequence for the break energy
$\delta_{GRB}h\nu'_0\sim200\,\xi$ keV, where
$\xi\sim$ 0.5--2 (Preece et al. 2000).
In our model, $\delta_{GRB}$ becomes $\sim2\gamma$.
Then, we obtain 
$\gamma\sim100\,\xi^{1/2}(h\nu_{DF}/5{\rm eV})^{-1/2}$.
On the other hand, we can derive 
$\nu_{DF}/\nu_{XRF}\sim(\theta_v/2)^2$,
which implies that we can also estimate the viewing angle.




\acknowledgments
We would like to thank the referee for useful comments and suggestions.
We are grateful to A.~Yoshida and T.~Murakami for helpful comments,
and W.~Naylor for a careful reading of the manuscript.
This work was supported in part by
Grant-in-Aid for Scientific Research 
of the Japanese Ministry of Education, Culture, Sports, Science
and Technology, No.00660 (KI), 
No.14047212 (TN), and No.14204024 (TN).

\clearpage

\begin{figure}
\plotone{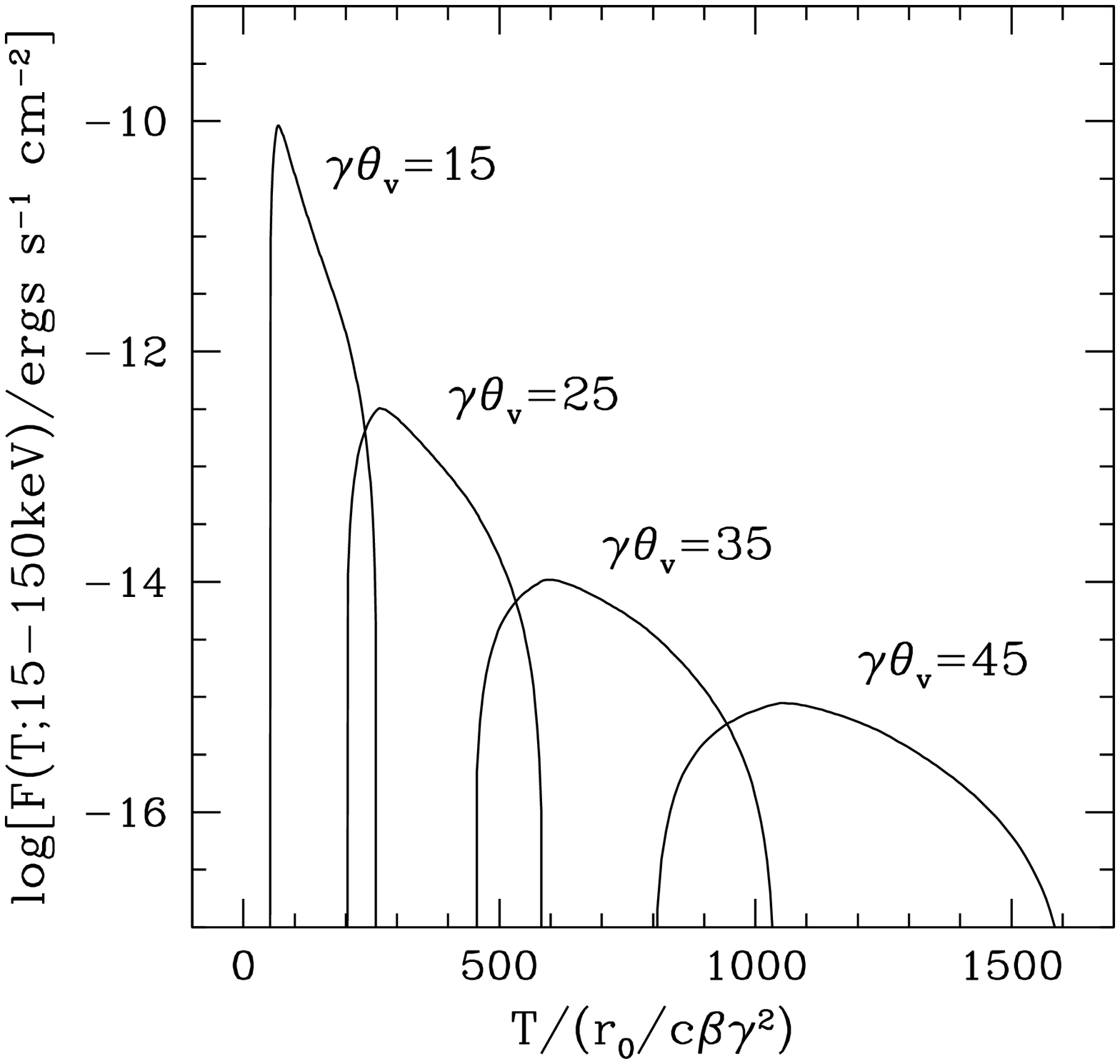}
\caption{
The light curves of the X-ray flash as a function of the 
normalized observed time $T/(r_0/c\beta\gamma^2)$,
where we adopt $r_0/\beta c\gamma^2=1$ sec.
We choose $\gamma\Delta\theta=5$, 
$\alpha_B=-1$, $\beta_B=-3$, 
$\gamma\nu'_0=200\,{\rm keV}$
and $D=1\,{\rm Gpc}$.
The flux is proportional to $D^{-2}$.
}
\label{fig:xrf}
\end{figure}

\begin{figure}
\plotone{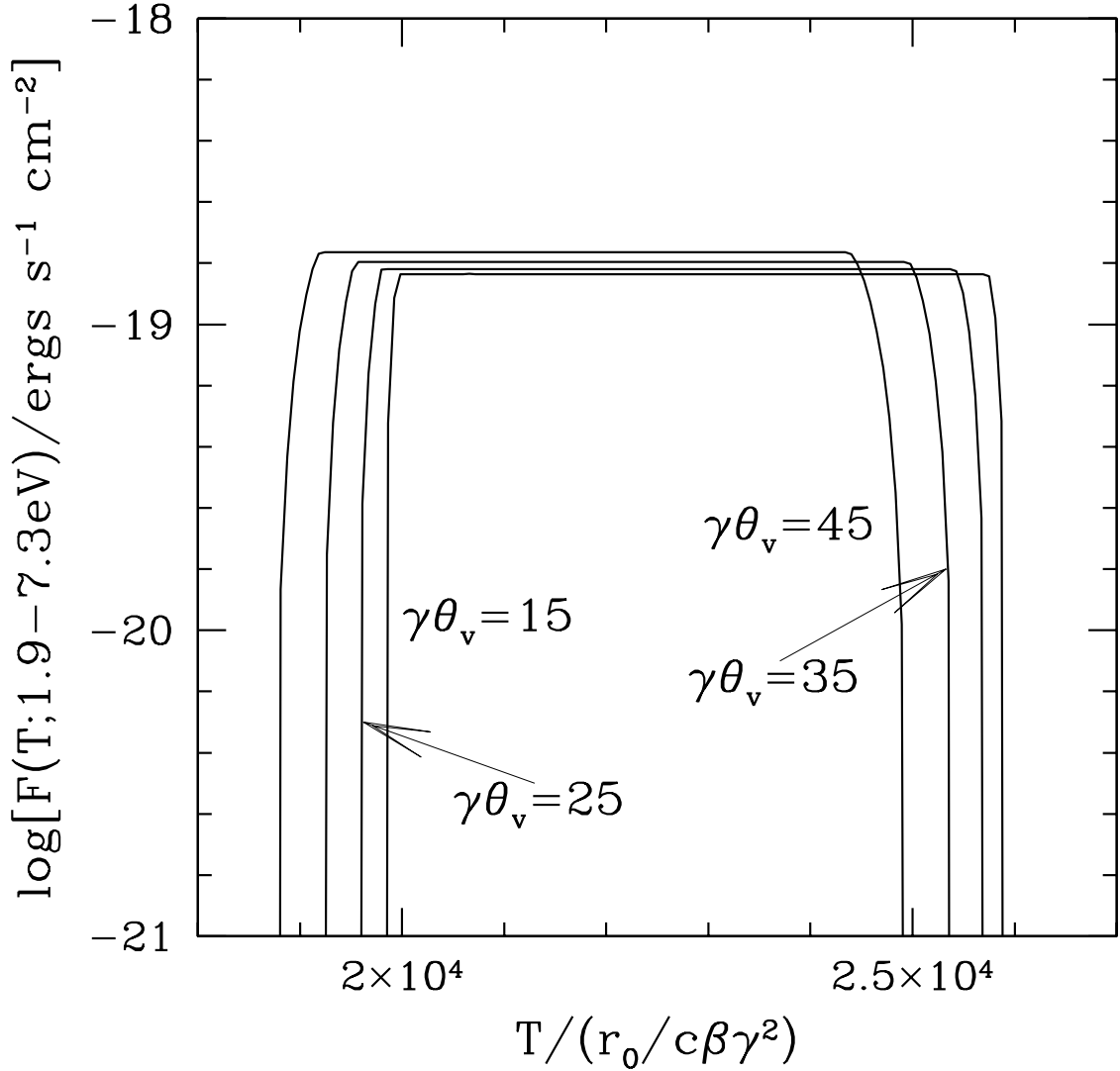}
\caption{
The light curves of the {\it delayed flash} as a function of the 
normalized observed time $T/(r_0/c\beta\gamma^2)$,
where we adopt $r_0/\beta c\gamma^2=1$ sec.
We choose $\gamma=100$, $\gamma\Delta\theta=5$, 
$\alpha_B=-1$, $\beta_B=-3$, 
$\gamma\nu'_0=200\,{\rm keV}$
and $D=1\,{\rm Gpc}$.
Our jet model predicts that 
the flux of the delayed flash
is almost constant ($F\sim 2\times10^{-19}D_{\rm Gpc}^{-2}{\rm ergs}
\ {\rm s}^{-1}{\rm cm}^{-2}$)
with both the observed time and the viewing angle.
}
\label{fig:df}
\end{figure}

\begin{figure}
\plotone{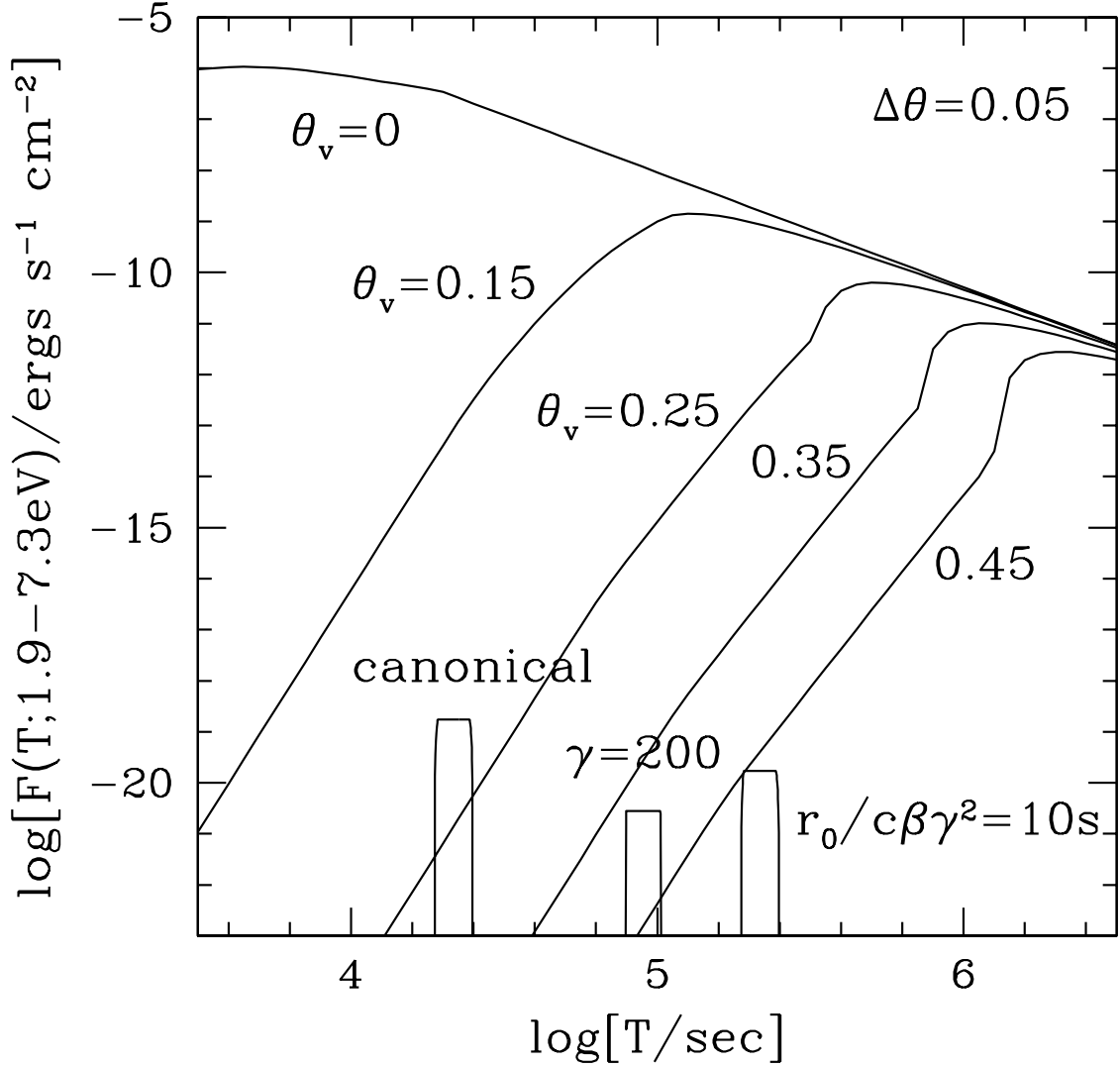}
\caption{
The light curves of the X-ray flash afterglow in the UV band are
shown by varying the viewing angle $\theta_v$.
We fix parameters as $\Delta\theta=0.05$,
$n=1\,{\rm cm}^{-3}$, $p=2.25$, 
$E=2\times10^{54}\,{\rm ergs}$,
$\varepsilon_e=0.1$, $\varepsilon_B=0.01$, 
and $D=1\,{\rm Gpc}$.
Boxes represent the light curves of the {\it delayed flash}
 in the same band. 
We choose a canonical set of parameters as $\gamma=100$ and
$r_0/\beta c\gamma^2=1\,{\rm sec}$.
The light curve of the delayed flash does not depend on
the viewing angle $\theta_v$ so much.
For comparison, we show the light curves of the delayed flash
with one of the fiducial parameters changed.
Note that all the flux is proportional to $D^{-2}$,
and the flux and the duration of the delayed flash are
proportional to $(r_0/\beta c\gamma^2)^{-1}$ and
$r_0/\beta c\gamma^2$, respectively.
}
\label{fig:ag1}
\end{figure}

\begin{figure}
\plotone{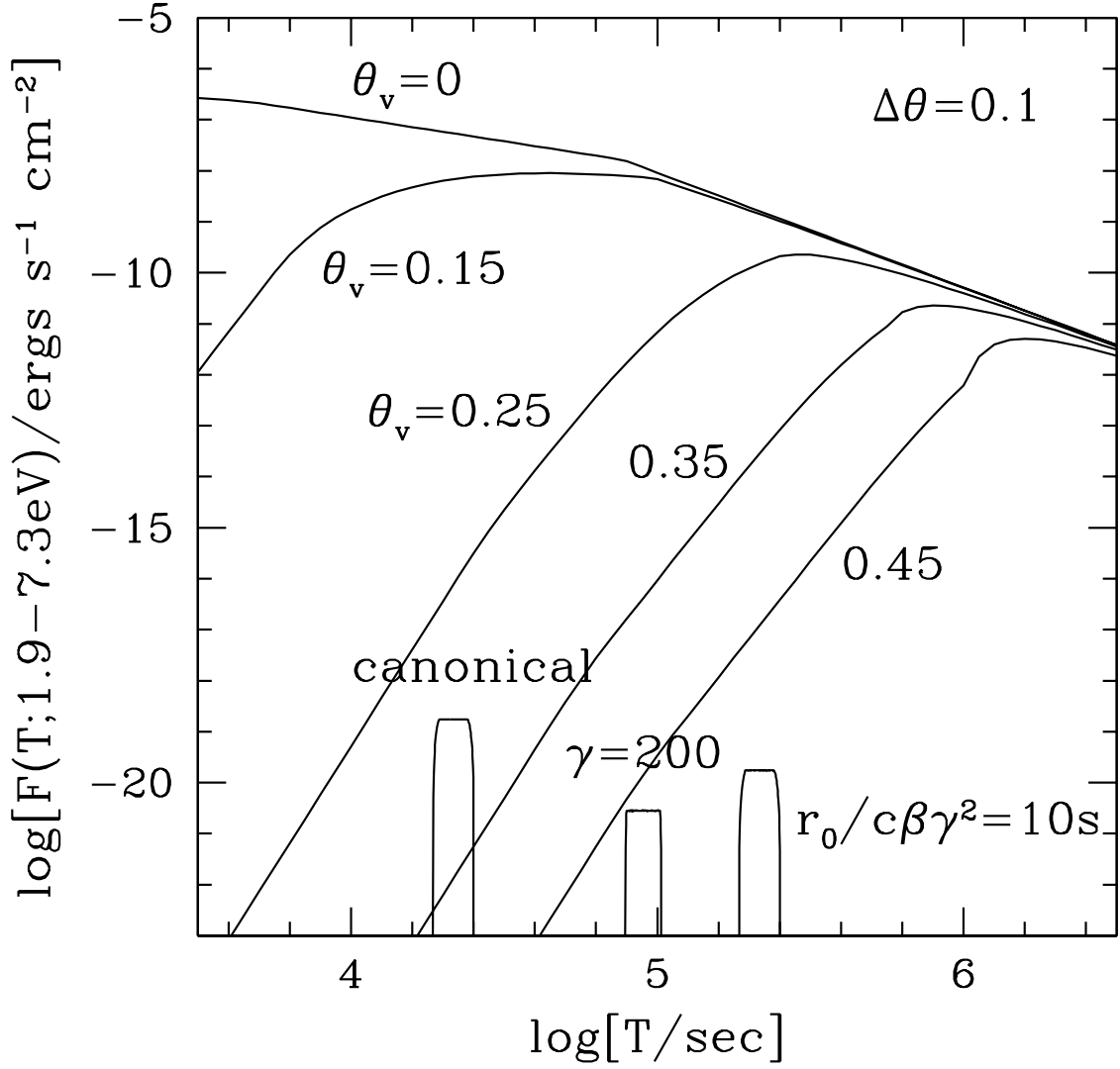}
\caption{
The same as Fig.~3 but for $\Delta\theta=0.1$ and 
$E=5\times10^{53}\,{\rm ergs}$. Note that the geometry corrected
total energy $(\Delta\theta)^2E/2$ is not altered.
}
\label{fig:ag2}
\end{figure}

\end{document}